\documentclass[conference]{IEEEtran}
\IEEEoverridecommandlockouts
\usepackage{cite}
\usepackage{amsmath,amssymb,amsfonts}
\usepackage{algorithmic}
\usepackage{algorithm}
\usepackage{graphicx}
\usepackage{textcomp}
\usepackage{xcolor}
\def\BibTeX{{\rm B\kern-.05em{\sc i\kern-.025em b}\kern-.08em
    T\kern-.1667em\lower.7ex\hbox{E}\kern-.125emX}}

\newif\ifanon
\anonfalse
\usepackage{booktabs}
\usepackage{url}
    
\begin{document}

\title{AI-Assisted GPU optimization of the Stochastic Variational Method for Few-Body Boson System }

\author{
\IEEEauthorblockN{Noritaka Shimizu\IEEEauthorrefmark{1},
                  Shigeyoshi Aoyama\IEEEauthorrefmark{2},
                  Yusuke Tsunoda\IEEEauthorrefmark{1},
                  Akira Nukada\IEEEauthorrefmark{1}}
\IEEEauthorblockA{\IEEEauthorrefmark{1}\textit{Center for Computational Sciences,
  University of Tsukuba}, Tsukuba, Japan\\
  \{shimizu, ytsunoda\}@nucl.ph.tsukuba.ac.jp, nukada@cs.tsukuba.ac.jp}
\IEEEauthorblockA{\IEEEauthorrefmark{2}\textit{Computing Research Center,
  High Energy Accelerator Research Organization (KEK)}, Tsukuba, Japan\\
  shigeyoshi.aoyama@kek.jp}
}


\maketitle

\begin{abstract}
The stochastic variational method (SVM) is one of the most powerful methods to solve quantum few-body systems precisely
in various fields, such as nuclear and atomic physics.
To the best of our knowledge, no SVM code optimized for GPUs has been reported.
We developed a parallel SVM code for few-body clusters of $^4$He atoms and optimized it for two GPU architectures, NVIDIA GH200 and AMD Instinct MI300A, with the entire code written by an AI coding agent.
On the algorithmic side, we introduce the secular-equation method  with the Gu--Eisenstat prescription for solving the generalized eigenvalue problem within the SVM framework.
The main part of the code tuning is to batch the many small matrices so that the GPUs are used efficiently, together with an array layout in memory optimized for coalesced addressing.
For the five-body system with 4800 SVM basis states, the tuned SVM code runs 16.2 (14.7) times faster on the NVIDIA GH200 (AMD MI300A) than the same tuned code on the Intel Xeon Max CPU system, and 168 (153) times faster than the first working CPU implementation. 
The achieved performance brings 6-body and larger cluster systems within reach.
\end{abstract}


\section{Introduction}

\footnote{(c) 2026 IEEE. Personal use of this material is permitted. Permission from IEEE must be obtained for all other uses, in any current or future media, including reprinting/republishing this material for advertising or
promotional purposes, creating new collective works, for resale or redistribution to servers or lists, or reuse of any copyrighted component of this work in other works.}

Quantum few-body problems, namely calculating ground-state and low-lying excited states of atomic, nuclear, and exotic clusters, are fundamental in computational physics. 
The Stochastic Variational Method (SVM) with a correlated Gaussian basis \cite{varga_prc,suzukivarga} is one of the most successful methods for obtaining high-precision solutions, particularly for problems with strong correlations, such as halo nuclei or weakly bound atomic clusters like helium dimers and trimers.
In the present work, we adopt the SVM to  evaluate a few-body system consisting of $^4$He atoms. 

The helium-4 cluster system has attracted much interest in connection with the Efimov physics.
Much effort has been devoted to elucidating this physics with the Gaussian expansion method \cite{hiyama2003gaussian}, density functional theories, and other approaches \cite{attobigio}.
Precise quantum many-body calculations beyond the five-body system remain difficult because they have to treat the strong short-range repulsion and are limited by available computational resources.

In the SVM, the wave function is written as a linear combination of correlated Gaussian basis states, which are selected stochastically from many candidates so that the variational energy is systematically lowered as the basis space is enlarged.
In the original SVM the selection is performed sequentially: candidates are generated and their energies evaluated one by one, and a candidate is adopted if a certain energy gain is achieved.
Such a sequential selection, however, hampers efficient parallel computation, and ``Parallel SVM'' was proposed as a parallel-computing-based extension of the SVM \cite{aoyama_npar}.
In this method a fixed number of candidate basis states are evaluated in parallel, which exploits the performance of modern massively parallel computers efficiently, and the best one is chosen among these candidates incrementally. 
In the present work, we follow this strategy which is expected to be beneficial also for GPU computing.

Most of the latest supercomputers are equipped with Graphics Processing Units (GPUs), which accelerate scientific computations by an order of magnitude over CPU systems.
Although the SVM algorithm is well suited to parallel computation, GPU programming remains a demanding task for physicists.
The recent growth of AI coding agents drastically reduces this burden: performance tuning is particularly well suited to them, because the correctness of every code modification can be verified automatically against reference results and the target is clearly defined, namely the measured wall time.

In the present work, we develop a highly tuned SVM code employing AI-coding and discuss its performance on GPU machines. 
This paper is organized as follows: 
the present Hamiltonian and the framework of the SVM are described in Sect.~\ref{sec:method},
Sect.~\ref{sec:code} is devoted to the algorithm and code improvements,
and the energy convergence and the code performance are discussed in Sect.~\ref{sec:result}.
Sect.~\ref{sec:summary} summarizes this work.

\section{Method}
\label{sec:method}

We present the Hamiltonian of the few-body $^4$He boson system and the SVM framework used to obtain its eigenstates.
Since the SVM is a variational method, the energy, namely the expectation value of the Hamiltonian, has to be evaluated efficiently.
The Hamiltonian is given in Subsect.~\ref{sec:hamiltonian},
and the SVM framework together with the formulae for evaluating the energy in Subsect.~\ref{svmframe}. The detailed formulae to compute the matrix elements are described in Subsect.~\ref{sec:me}.
The AI-assisted coding process is described in Subsect.~\ref{sec:ai-coding}.

\subsection{Hamiltonian of $^4$He atomic cluster}
\label{sec:hamiltonian}

We begin with the Hamiltonian of $N$-body system, including the two-body potential as 
\begin{equation}
    H = T + V = -\sum_{a=1}^{N_J}  \frac{\hbar^2}{2m_a} \nabla_a^2
    + \sum_{i<j}^N V(r_{ij}), 
\end{equation}
where $\nabla_a^2$ and $m_a$ are the Laplacian with respect to the Jacobi coordinate $a$ and its reduced mass, respectively. The symbol $N_J$ is the number of the Jacobi coordinates, namely $N_J=N-1$. 
The symbol $r_{ij}$ is the relative distance between two particles $i$ and $j$.
The mass of the $^4$He atom, $m$, 
is taken as $\hbar^2/(mk_B) = 12.12$ K\AA$^2$ \cite{pang2020breakup}.
In the present work, the two-body potential is approximated as a two-range Gaussian function of the relative distance $r$ as 
\begin{eqnarray}
    V(r) &=& \sum_{\nu=1}^2 v_\nu\, e^{-(r/\mu_\nu)^2} 
\end{eqnarray}
with $v_1 =2000$ K, $v_2=-8.28233$ K, $\mu_1 =1$\AA, and $\mu_2=3.48813$ \AA.  
These parameters were determined to reproduce the scattering length and the dimer energy of the LM2M2 potential \cite{lm2m2}, and this potential was adopted in some previous studies \cite{yoshida_hpcasia,aoyama_iccsa2026}. 
The first term represents a strong repulsive force caused by the Pauli principle, and the second a shallow attractive van der Waals force.
This strong repulsion makes it difficult for many other methods to compute the wave function precisely. 

\subsection{Stochastic Variational Method}
\label{svmframe}

\begin{figure}[htbp]
\centerline{\includegraphics[width=.45\columnwidth]{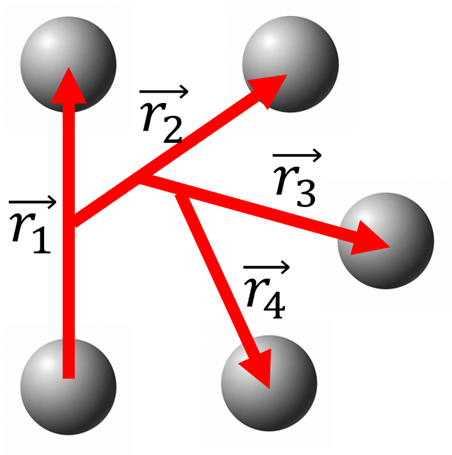}}
\caption{Jacobi coordinates, $\vec{r}_a$, of five-body system ($N=5$ and $N_J=4$).}
\label{fig:jacobi}
\end{figure}

The SVM wave function is a linear combination of the correlated Gaussian basis states as
\begin{equation}
    \Psi_{N_b}({\bf x}) =\sum_{k=1}^{N_b} c_k  {\cal S} \phi_{A_k}({\bf x}), 
    \label{eq:wf}
\end{equation}
where ${\cal S}$ is the symmetrization operator for the Bose statistics.
The symbol $\mathbf{x}=( \vec{r}_1, \vec{r}_2, \cdots, \vec{r}_{N_J})$ is 
a set of Jacobi coordinates for an $N$ particle system as shown in Fig.~\ref{fig:jacobi} with the number of the coordinates being $N_J=N-1$.
The Jacobi coordinates are defined as relative coordinates of the particles so that the center-of-mass motion is subtracted exactly.
Only one of the possible sets is shown; any rearrangement of the Jacobi coordinates is absorbed into the matrix $A$ introduced below.
In the present work we consider only the $J^\pi=0^+$ state, so that the global vector for the angular momentum is not required and the functional form is kept simple \cite{suzukivarga}.

The correlated Gaussian basis $ \phi(\bf{x})$ is defined as
\begin{equation}
    \phi_A(\textbf{x}) = \exp\left( -\frac12 \textbf{x}^T A \textbf{x} \right) = \exp\left( -\frac12 \sum_{a,b=1}^{N_J} A_{ab}\,\vec{r}_a\cdot\vec{r}_b \right) ,  
\end{equation}
where the symbol $A$ is a symmetric positive definite matrix of size $N_J\times N_J$ and thus contains $N_J(N_J+1)/2$ parameters for each matrix. 
In the SVM framework, the matrices $A$ are generated randomly, and the best one to minimize the energy is selected based on the variational principle.

The eigenenergies and the coefficients $c_k$ in Eq.~(\ref{eq:wf}) are determined by solving the Hill--Wheeler equation, or the generalized eigenvalue problem, written here in the bra-ket notation with $\phi_A(\textbf{x}) = \langle \textbf{x}|A \rangle$, as 
\begin{equation}
    \sum_{k'} H_{kk'} c_{k'}  
    = E \sum_{k'}N_{kk'}  c_{k'}.
    \label{eq:gev}
\end{equation}
\begin{equation}
    H_{kk'} = \langle A_k |H {\cal S}|A_{k'}\rangle, \ \ \  
    N_{kk'} = \langle A_k |{\cal S}| A_{k'} \rangle .
    \label{eq:mele}
\end{equation}
Thus, the wave function is written as a linear combination of the symmetrized correlated Gaussian basis, whose matrix elements can be computed efficiently employing the Gaussian integral as discussed in Subsect.~\ref{sec:me}. 

Algorithm~\ref{alg:svm} shows the parallel SVM. At each step, where $K$ basis states have been accepted, $N_t$ candidates are evaluated in parallel and the one that lowers the energy most is accepted.
\begin{algorithm}
\caption{Algorithm of the parallel SVM \cite{aoyama_npar}.}
\label{alg:svm}
\begin{algorithmic}[1]
\FOR{$K=0,1,...,N_b-1$}
  \STATE prepare $N_t$ random matrices $A^{(1)},\dots,A^{(N_t)}$
  \FOR{$t = 1,\dots,N_t$}
   \STATE set $A_{K+1}=A^{(t)}$
    \STATE compute $H_{1:K+1,K+1}$, $N_{1:K+1,K+1}$  
    \STATE solve Eq.~(\ref{eq:gev}) with dimension $K+1$ $\rightarrow$ energy $E^{(t)}$
  \ENDFOR
  \STATE $t \leftarrow \arg\min E^{(t)}$
  \STATE $A_{K+1}\leftarrow A^{(t)}$ and reconstruct $H$ and $N$
\ENDFOR
\end{algorithmic}
\end{algorithm}

\subsection{Matrix elements in SVM}
\label{sec:me}

The most time-consuming part of the SVM method is to evaluate the Hamiltonian and norm matrix elements in Eq.~(\ref{eq:mele}).
For identical Bose particles, it is necessary to symmetrize the wave function under the exchange of particles.
A particle permutation acts on Jacobi
coordinates as an orthogonal matrix $P$, and transforms a Gaussian into another Gaussian with matrix $P^TAP$. The symmetrized basis element is
\begin{equation}
 \langle A_k |H{\cal S} |A_{k'} \rangle
 = \sum_{p} \langle A_k | H |A_{k'}^{(p)} \rangle, 
 \label{eq:perm}
\end{equation}
where $A_{k'}^{(p)} = P_p^{T} A_{k'} P_p$ and $P_p$ is an $N_J \times N_J$ matrix representing the exchange of particles corresponding to the permutation $p$. 
Since the Hamiltonian commutes with ${\cal S}$,
it is sufficient to symmetrize the wave function on one side.
Consequently, the computation of every matrix element demands a sum over the $N!$ permutations, which makes the computational cost scale as $O(N!)$.
The norm matrix element $N_{kk'}$ is computed in a similar way.

The overlap between two different basis states is calculated efficiently by employing the Gaussian integral as
\begin{equation}
    \langle A_k | A_{k'}^{(p)} \rangle =  
    \left( \frac{(2\pi)^{N_J} }{\det(B)}
    \right)^{3/2}, \ \ B=A_k + A_{k'}^{(p)}
\end{equation}
where the determinant is calculated by the Cholesky decomposition of the matrix $B$. The norm matrix element is calculated as the summation concerning $p$.
Hereafter, we omit the index $(p)$ for simplicity.

The matrix elements of the kinetic term are computed as
\begin{eqnarray}
    \label{eq:kine_mat}
    \langle A_k | T |A_{k'}\rangle
    &=& \frac{3}{2}\,\langle A_k |A_{k'}\rangle\,
        \textrm{Tr}\left( A_k B^{-1} A_{k'} M \right),
\end{eqnarray}
where $M$ is the diagonal matrix whose elements are $M_{ab}=\hbar^2\delta_{ab}/m_a$ and $m_a$ is the reduced mass of the Jacobi coordinate $a$.
Thus, the computation cost of Eq.~(\ref{eq:kine_mat}) is 
$O(N_J^3)$ due to the Cholesky decomposition of $B$ to construct $B^{-1}$.

For the potential term, the relative position vector $\vec{r}_{ij}$ of the particle pair $(i,j)$ is written as a linear combination of the Jacobi coordinates as 
\begin{eqnarray}
  \vec{r}_{ij} &=& \sum_{a=1}^{N_J} w^{(i,j)}_a\,\vec{r}_a 
  \nonumber \\
  r_{ij}^2 &=& \mathbf{x}^T  
  \mathbf{w}^{(i,j)}\,(\mathbf{w}^{(i,j)})^{T} \mathbf{x},
  \label{eq:wdef}
\end{eqnarray}
where the constant coefficient vector $\mathbf{w}^{(i,j)}\in\mathbb{R}^{N_J}$ is given by the
inverse Jacobi transformation. Thus, the potential is rewritten in terms of the Jacobi coordinates as 
\begin{eqnarray}
 V(r_{ij}) = \sum_\nu v_\nu \exp(-\mathbf{x}^T 
 \mathbf{w}^{(i,j)}\,(\mathbf{w}^{(i,j)})^{T}
 \mathbf{x} / \mu_\nu^2) .
\end{eqnarray}
Thanks to the matrix determinant lemma, 
the matrix elements of the potential term are calculated efficiently as 
\begin{eqnarray}
\label{eq:vmat}
    \langle A_k | V |A_{k'}\rangle
    &=& \langle A_k |A_{k'}\rangle \sum_{\nu,i,j} 
     \frac{v_\nu}{(1+2s^{(ij)}/\mu_\nu^2)^{3/2}}.     \nonumber
     \\
     s^{(ij)} &=& (\mathbf{w}^{(i,j)})^{T}B^{-1}\mathbf{w}^{(i,j)}.
\end{eqnarray}
While $ s^{(ij)}$ is independent of $\nu$, it depends on $k, k', p, i,$ and $j$. 
The pair index $(i,j)$ runs over $N(N-1)/2$ values, so that the computational cost per matrix element is $O(N^4)$, 
which is heavier than the kinetic term in Eq.~(\ref{eq:kine_mat}) 
and is the bottleneck part of the whole SVM computation.

A typical SVM calculation builds the variational basis incrementally: at each step, hundreds to thousands of trial Gaussian basis states are evaluated, and the best one, which has the lowest energy eigenvalue, is added to the basis. 
For every candidate, the generalized eigenvalue problem in Eq. (\ref{eq:gev}) is solved to obtain the energy. 
For physically interesting problems the basis size reaches $N_b = 100$--$10000$ or more, depending on the number of particles, and the costs of the matrix-element evaluation and solving the eigenvalue problems become prohibitive on a single CPU.
The former cost can be reduced through GPU-oriented code optimization, while the latter can be reduced by introducing the secular-equation method.
These improvements are described in the next section.

\subsection{AI-assisted coding}
\label{sec:ai-coding}

By using AI agents, we developed the whole SVM code from scratch. 
We gave many instructions to the AI agents (Claude Opus mainly) with natural language, the code was developed gradually in a few months. 
We do not touch the code directly except for a few trivial modifications. 
Concerning the GPU optimization, we asked the AI agent to measure the profile data, and it measured and analyzed the data suggesting us how to optimize the code. 
At the early stage, the code was developed by the one of the authors (N.S.) who has little experience in the CUDA developments, and wrote the "baseline" code including the technique discussed in Subsect.~\ref{sec:secular}. 
After that, the discussions with High-Performance-Computing experts accelerates the CUDA optimization, indicating that the domain knowledge of the CUDA coding and GPU hardware is essential even for AI-coding. 
The code development was completed after more than 1000 brief interactions with AI and many trial-and-errors including the technique discussed in Subsect.~\ref{sec:soa}. 
The production code consist of 13,682 lines of Fortran, CUDA and HIP codes.

\section{Performance improvements}
\label{sec:code}

In this section, we focus on two improvements which dominate the gain over a straightforward implementation, and they act on different components. The secular-equation treatment of the
generalized eigenvalue problem removes the $O(N_b^3)$ cost from the
per-candidate loop, which is discussed in Subsect.~\ref{sec:secular}.
The transposition of the memory layout makes the matrix-element kernel use the memory system efficiently, which is discussed in Subsect.~\ref{sec:soa}.

The computations for $N_t$ candidates are mutually independent, so that the $N_t$ candidates are distributed evenly over the $N_{\rm proc}$ MPI processes, each of which handles $N_t/N_{\rm proc}$ candidates. 
Within a process
the workload is divided between GPU and CPU: the matrix elements are
computed on GPU when available, while the eigenvalue problems of the
candidates are solved on the CPU with OpenMP threads. The candidates of a
process are further split into chunks of the permutations, 
and the chunks are pipelined so that the CPU-side processing of one chunk overlaps with the GPU execution of the following ones.

\subsection{Solving the generalized eigenvalue problem with the secular equation}
\label{sec:secular}

At each SVM step, a $(K+1)$-dimensional generalized
eigenvalue problem must be solved for each of the $N_t$ candidates.
Solving it from scratch costs $O(K^3)$, so that this stage alone would cost
$O(N_t K^3)$ per step and $O(N_t N_b^4)$ for the whole run, which overtakes
the cost of the matrix elements as soon as $N_b$ becomes large,
even though its efficient GPU implementation is feasible \cite{tokura}.

In Ref.~\cite{suzukivarga}, the secular-equation technique was introduced in the SVM to avoid this problem.
Let the eigenpairs of the $K$-dimensional problem be already solved, with the
eigenvectors $\hat{U}$ normalized with respect to the norm matrix,
$\hat{U}^{T} N_K \hat{U} = I$ and
$\hat{U}^{T} H_K \hat{U}= \mathrm{diag}(\varepsilon_i)$.
Appending a candidate as the $(K+1)$-th basis state adds one row and one
column to the $K\times K$ Hamiltonian and norm matrices.
Transforming the new row and column into the eigenbasis of the $K\times K$ problem gives
\begin{equation}
  v_k = \sum_{k'=1}^{K}\hat{U}_{k'k}\,H_{k',K+1}, \qquad u_k = \sum_{k'=1}^{K}\hat{U}_{k'k}\,N_{k',K+1}
  \label{eq:u_hn}
\end{equation}
for $k=1,\dots,K$, together with the diagonal elements $H_{K+1,K+1}$ and $N_{K+1,K+1}$.
By applying the Gram-Schmidt orthogonalization so that the new basis state is orthogonal to the fixed $K$ basis states, the generalized eigenvalue problem is rewritten as a standard eigenvalue problem of the symmetric \emph{arrowhead}-structure matrix
\begin{equation}
  \widehat{H} =
  \begin{pmatrix}
    \varepsilon_1 &        &               & \beta_1    \\
    & \ddots &               & \vdots \\
    &        & \varepsilon_K & \beta_K \\
    \beta_1  & \cdots & \beta_K  & \alpha
  \end{pmatrix},
  \label{eq:arrow}
\end{equation}
with
$\gamma = N_{K+1,K+1} - \mathbf{u}^{T}\mathbf{u}$ and
\begin{equation}
  \alpha = \frac{H_{K+1,K+1} - 2\,\mathbf{u}\!\cdot\!\mathbf{v}
  + \sum_k \varepsilon_k u_k^2}{\gamma},
  \qquad
  \beta_k = \frac{v_k-\varepsilon_k u_k}{\sqrt{\gamma}},
  \label{eq:arrowpar}
\end{equation}
where the empty matrix elements are zero.
The $K+1$ eigenvalues of Eq.~(\ref{eq:arrow}) are the roots of the secular
equation $\det(\hat{H}-EI)=0$, namely 
\begin{equation}
  f(E) = (\alpha - E) - \sum_{k=1}^{K}\frac{\beta_k^2}{\varepsilon_k - E} = 0 .
  \label{eq:sec}
\end{equation}
If $\gamma$ is too small or non-positive,  the candidate is linearly dependent on the accepted basis, and the candidate is rejected.
Thanks to Cauchy's interlacing theorem, the roots interlace the eigenvalues $\varepsilon_k$, so that each of them is bracketed in
advance and is obtained by the bisection method in a numerically stable manner.

In the candidate loop only a few of the lowest eigenvalues are required.
The basis transformation of $\hat{U}$ costs $O(K^2)$ and each evaluation of $f(E)$ costs $O(K)$, so that the candidate energy is evaluated in $O(K^2)$ instead of $O(K^3)$, and no eigenvector is computed at this stage. While this $\hat{U}$ transformation is achieved as a matrix-vector product for each candidate, we collect the vectors of many candidates into a chunk and perform the products as a matrix-matrix product for further acceleration.
This is the ``Secular equation'' component of Fig.~\ref{fig:breakdown}, which is executed $N_t$ times per step.

The eigenvectors are needed only once per step, for the candidate that is actually accepted, because the $(K+1)$-th step requires all the eigenpairs of the $K\times K$ problem.
They are constructed from the roots of Eq.~(\ref{eq:sec}) by the Gu--Eisenstat prescription \cite{gu_eisenstat}: building the eigenvectors
directly from the $\beta_k$ loses their orthogonality when two roots lie close to
each other, and the prescription instead reconstructs the couplings from the
computed roots, which restores orthogonality to working accuracy.
The cost of this update is $O(K^3)$, but it is dominated by a single matrix--matrix product, the back transformation of the eigenvectors into the
original basis, which runs far faster than a general eigenvalue solver such as \texttt{dsygvd}.
This is the ``Incremental update'' component, executed once per step.
This step is parallelized using CPU OpenMP threads, with GPU acceleration enabled for $K\geq 2000$.

Since each incremental update uses the eigenpairs of the preceding step as exact input, rounding errors accumulate over the steps.
We therefore reconstruct the eigenpairs from scratch every 100 basis states, which is called ``Eigensolver'' component in Tab.~\ref{tab:results} and is executed 48 times in a run with $N_b=4800$ performed on GPU (\texttt{cusolverDnDsygvd} / \texttt{hipsolverDsygvd}) if available.

The cost per step is thus $N_t \times O(K^2) + O(K^3)$ instead of $N_t \times O(K^3)$.
The benchmark of Fig.~\ref{fig:breakdown} shows the effect of this improvement: on the GH200 the per-candidate part costs 10.1~s although it is executed $N_t=1200$ times more often than the once-per-step update, which costs 14.6~s, while the 48 library solves cost only 3.2~s.
The MPI communication follows the same division. Only the row of the winning candidate have to be exchanged, which is $O(K)$ data per step;
the $K\times K$ eigenvector matrix is replicated on every rank and updated
redundantly, so that no $O(K^2)$ matrix is ever communicated. The measured ``MPI synchronization'' is accordingly at the percent level, 1.2~s out of 70.8~s.

\subsection{Memory layout of the matrix-element kernel}
\label{sec:soa}

The matrix-element computation of the $K$-th basis decomposes into $N! \times K \times N_t$ mutually
independent tasks, each of which builds an $N_J\times N_J$ matrix $B$,
factorizes it by the Cholesky decomposition, and computes the matrix elements in Eqs.~(\ref{eq:kine_mat}) and (\ref{eq:vmat}).
With the $O(N^4)$ cost of one such task and the $N_b(N_b+1)/2$ matrix elements
of the whole run, the total cost is $O(N^4 N! N_b^2 N_t)$.
For the five-body system with $N_t=1,200$ already $N!N_t = 144,000$ tasks are
available even at $N_b=1$, and thus the computation is embarrassingly parallel and maps
naturally onto a GPU: one thread block is assigned to a candidate basis and its threads stride over the permutations, that is, \emph{one thread
processes one permutation}.
Since the number of particles is fixed through a whole computation, the kernel is
instantiated as a template over $N_J=N-1$, all inner loops are unrolled, and every intermediate matrix is held in registers.

The permuted matrices $A_{k'}^{(p)} = P_p^{T}A_{k'}P_p$ do not depend on the
index of the accepted basis state, so they are computed once per candidate and reused for all $N_b$ pairs. 
How they are stored turns out to be decisive.
In the natural array-of-structures (AoS) layout, the $N_J^2$ elements of one
permutation are contiguous, so that adjacent threads, which handle adjacent
permutations, read addresses $N_J^2$ doubles apart. Profiling with NVIDIA Nsight Compute showed that each 32-byte memory sector then delivered only 8
useful bytes, that the L1 cache throughput was the saturated resource at
67\%, and that the DRAM utilization was negligible at 0.02\%: the kernel was
limited by the number of memory transactions, not by the bandwidth.

The remedy is to transpose the array indices into a structure-of-arrays (SoA)
layout in which the permutation index is the innermost and contiguous one. 
Namely, the matrix $A_{i,j}^{(p)}$ is stored as $A_{1,1}^{(1)},A_{1,1}^{(2)},A_{1,1}^{(3)}  \cdots$, so that adjacent threads read adjacent doubles and the loads of a warp coalesce
into full sectors. The arithmetic is untouched, so that the energies are
bit-identical to those of the previous layout, while the matrix-element stage
becomes 1.3--1.6 times faster.

The same transformation applies to the CPU code, where the vector lanes play the role of the threads. 
Vectorizing inside one small matrix fails, because the innermost contraction has length $N_J=4$ and fills only half of the eight lanes of a 512-bit register, and because the Cholesky factorization is sequential within one matrix. 
Instead the permutation loop is moved from the
outermost to the innermost position and tiled, with the same structure-of-arrays layout, so that each vector instruction advances the same
elementary step of eight independent factorizations. Non-positive pivots
cannot break out of a vector loop and are handled by lane masking, the
software counterpart of the divergence handling of the GPU.
The correspondence is one to one --- GPU thread $=$ SIMD lane, coalesced
access $=$ unit stride --- and reflects a single requirement, namely that the
parallel dimension be contiguous in memory.

\section{Results}
\label{sec:result}
We show the energy convergence of the SVM method in Subsect.~\ref{sec:Npar}, and the code performance in Subsect.~\ref{sec:perf}.
A brief discussion of the 6-body system is shown in Subsect.~\ref{sec:6body}.

\subsection{Convergence of the SVM method.}
\label{sec:Npar}

\begin{figure}[htbp]
\centerline{\includegraphics[width=.9\columnwidth]{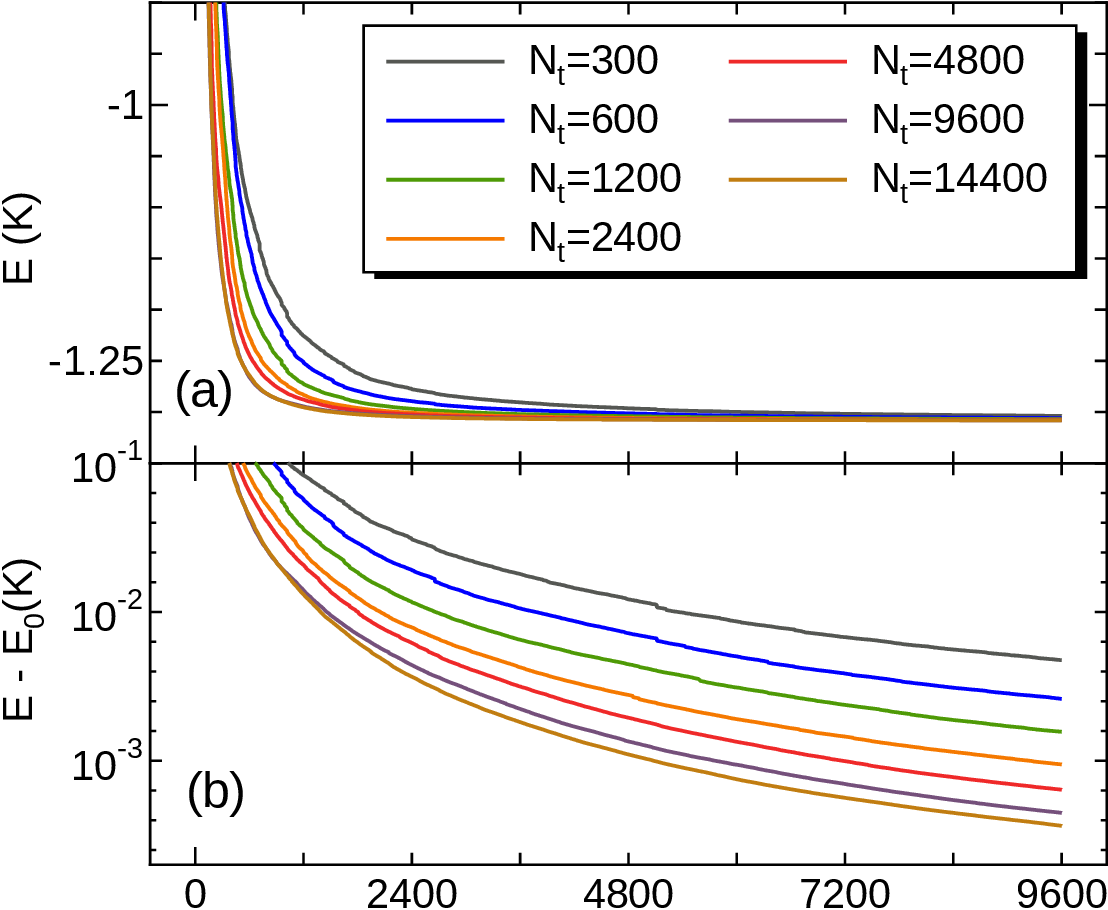}}
\caption{(a) Convergence of the ground-state energies of the five-body system against the number of the basis states $N_b$.
The grey, blue, green, orange, red, purple and brown lines denote the SVM results with $N_t=300, 600, 1200, 2400, 4800, 9600,$ and $14400$, respectively.  (b) Same as (a), but the vertical axis shows the energy subtracted by the estimated exact energy, $\Delta E = E - E_0$, on a logarithmic scale.}
\label{fig:Npar}
\end{figure}

As an example, we show the energy convergence of the five-body system against the number of the basis states $N_b$ in Fig.~\ref{fig:Npar} (a). As $N_b$ increases, the energy decreases monotonically and is nearly converged by $N_b \sim 4800$ for any $N_t$, although the converged value depends slightly on $N_t$.
Figure \ref{fig:Npar} (b) shows the energy difference $\Delta E$ from the estimated exact energy on a logarithmic scale. 
The exact value is estimated as $E_0=-1.30851(1)$ K 
employing the energy-variance extrapolation \cite{imadakashima}, in which the energy is extrapolated to zero energy variance.
The details of this extrapolation method will be reported elsewhere. 
The energy decreases with $N_b$, and the convergence behavior improves as $N_t$ increases.
The gain becomes progressively smaller: that from $N_t=9600$ to $N_t=14400$ is only 0.09~mK, indicating convergence with respect to $N_t$.

  \begin{table}[htbp]
    \centering
    \caption{SVM energies of the five-body $^4$He cluster with respect to
    the number of trial bases per basis function, $N_t$.
    Runs were performed with the tuned code and  $N_b=9600$ on 4 nodes of Miyabi-G.}
    \label{tab:conv_ntrial}
    \begin{tabular}{rlrr}
      \toprule
      $N_t$ & \ \ \ $E$ (K) & $\Delta E$ (mK) & Wall time (s) \\
      \midrule
        300 & $-1.303765$ & $4.74$ &  231.8 \\
        600 & $-1.305901$ & $2.60$ &  277.8 \\
       1200 & $-1.306942$ & $1.57$ &  381.7 \\
       2400 & $-1.307563$ & $0.95$ &  577.1 \\
       4800 & $-1.307869$ & $0.64$ &  987.3 \\
       9600 & $-1.308062$ & $0.45$ & 1782.7 \\
      14400 & $-1.308144$ & $0.36$ & 2574.2 \\
      $E_0$ (est.) & $-1.30851(1)$ & & \\
      \bottomrule
    \end{tabular}
    \\[2pt]
    \footnotesize
    The exact energy, $E_0$, is estimated with the energy-variance extrapolation.
    $\Delta E$ is the energy relative to $E_0$. 
  \end{table}
Table~\ref{tab:conv_ntrial} lists the wall time as a function of $N_t$.
It is approximately linear in $N_t$ as 
\begin{equation}
   \textrm{Wall time} = 181.5 + 0.1664\,N_t.
   \label{eq:walltime}
\end{equation}
The constant term is the
$N_t$-independent cost (eigensolver, incremental update and I/O),
while the slope reflects the matrix-element evaluation, which scales linearly with the number of trials.
Since the computational cost also scales as $N_b^2$, $N_t$ and $N_b$ have to be chosen as a trade-off under a given wall-time budget.

\subsection{Code Performance}
\label{sec:perf}

We present the wall time of the tuned SVM code for the five-body system with MPI parallelization.
The SVM calculations were performed with $N_b=4800$ and $N_t=1200$ and their wall times were measured.
Four devices are used on every platform, namely 4 nodes for Miyabi-G, 2 nodes (4 CPUs) for Miyabi-C, and 1 node (4 APUs) for Sirius supercomputer, which are summarized in Tab.~\ref{tab:platforms}.

\begin{table*}[t]
  \centering
  \caption{Computing platforms for the benchmark test.}
  \label{tab:platforms}
  \small
  \begin{tabular}{@{}llll@{}}
    \toprule
     & Miyabi-C (CPU) & Miyabi-G (GPU) & Sirius (GPU) \\
    \midrule
    Processor            & Intel Xeon Max 9480            & NVIDIA GH200 (Grace 72c + H100) & AMD Instinct MI300A \\
    Devices used         & 4 CPUs (224 cores)          & 4 GH200                         & 4 APUs \\
    FP64 peak / device   & 3.4 TFLOP/s        & 34 TFLOP/s (vector)       & 61.3 TFLOP/s (vector) \\
    Compilers            & ifort 2021.10.0 + MKL                    & nvcc CUDA 13.0, nvfortran     & ROCm 7.2.0 + gfortran 11.5.0  \\
    Numerical libraries  & MKL    2023.2.0            & cuBLAS / cuSOLVER               & hipBLAS / rocSOLVER \\
    Parallelization      & 4 MPI $\times$ 56 OpenMP       & 4 MPI $\times$ 72 OpenMP        & 4 MPI $\times$ 23 OpenMP$^a$ \\
    Mem BW / device & 1.6 TB/s  & 4.0 TB/s  & 5.3 TB/s \\ 
\bottomrule
  \end{tabular}
  \\
  $^{\mathrm{a}}$ Although the APU contains 24 CPU cores, 23 OpenMP threads are invoked to avoid OS noise. 
\end{table*}

\begin{figure}[htbp]
\centerline{\includegraphics[width=1.\columnwidth]{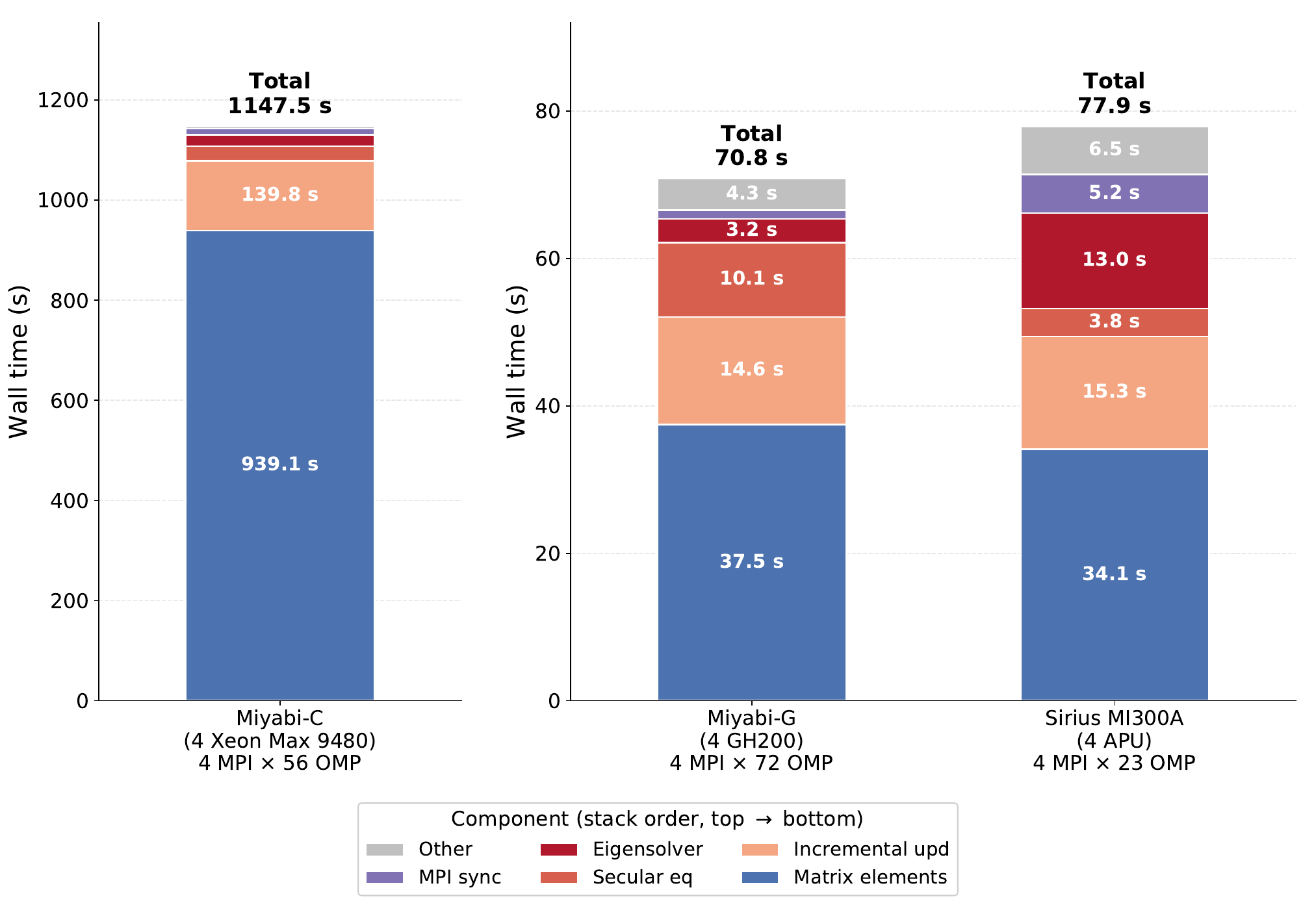}}
\caption{Wall-time breakdown of the tuned SVM code on Miyabi-C with Intel Xeon CPU (left), Miyabi-G with GH200 (middle), and Sirius with MI300A (right). 
The wall time of the SVM computation of the five-body system with $N_b=4800$ and $N_t=1200$ for 4 devices with MPI parallel.
The components are the matrix-element computation (blue), the incremental update (orange), the secular equation (light red), the eigensolver (dark red), the MPI synchronization (purple), and the others (grey). Note that the vertical scale of the left panel differs from that of the other two.}
\label{fig:breakdown}
\end{figure}

Table~\ref{tab:results} shows the wall time decomposed into six components, corresponding to the steps of the SVM algorithm described in Sect.~\ref{sec:method}.
The times are measured on the three devices with AI-tuned code and with the first-implemented code (baseline) as a comparison. 
The performance tuning on the CPU substantially accelerates the code, achieving a speedup of up to a factor of 10; the port to the GH200 reaches 168.
Figure~\ref{fig:breakdown} also shows the measured time as breakdown bars. 
``Matrix elements'' is the evaluation of the couplings of every candidate to
the accepted basis states, Eqs.~(\ref{eq:kine_mat}) and (\ref{eq:vmat}),
each of which carries the sum over the $N!$ permutations.
``Secular equation'' is the energy evaluation of each candidate from those couplings, ``Incremental update'' is the update of all eigenpairs by the secular-equation method once the
winning candidate has been accepted, and ``Eigensolver'' denotes the full solution of the generalized eigenvalue problem using the \texttt{dsygvd} library call or its variants.  
``MPI synchronization'' is the collective communication that selects the
globally best candidate, and ``Other'' collects the remaining host-side work
such as the candidate generation and the file I/O.

On the Sirius supercomputer we found large synchronization delays among the MPI processes when 24 OpenMP threads were used per APU, even though each APU has 24 CPU cores.
Running with 23 threads per APU gives better performance and more stable wall times than with 24 threads.


\begin{table*}[t]
  \centering
  \caption{Wall-time breakdown for the five-body SVM run
    ($K=4800$ basis states, $N_t=1200$ trials per step; four devices on
    every platform; seconds). ``CPU baseline'' is the pre-optimization
    code with the original flat-MPI decomposition
    (112 MPI $\times$ 2 OpenMP); the tuned columns use the
    configurations of Table~\ref{tab:platforms}. The CPU-to-GPU step
    combines the port and the hardware: the software-only gain on
    identical CPU hardware is the $10.4\times$ of the second column.
    ``Eigensolver'' is the full generalized eigensolve refreshed every
    100 basis steps (48 calls): QR-based \texttt{dsygv} in the CPU
    baseline, divide-and-conquer \texttt{dsygvd}
    (MKL / cuSOLVER / rocSOLVER) in the tuned columns; the
    per-step rank-one update is the separate ``incremental update'' row.}
  \label{tab:results}
  \small
  \begin{tabular}{@{}lrrrr@{}}
    \toprule
    Component & CPU baseline & CPU tuned & GH200 tuned & MI300A tuned \\
    \midrule
    Matrix elements      & 7435.9 & 939.1 & 37.5 & 34.1 \\
    Incremental update   & 3217.0 & 139.8 & 14.6 & 15.3 \\
    Secular equation     &   52.4 &  29.0 & 10.1 &  3.8 \\
    Eigensolver (full solve)      & 417.3 &  22.6 &  3.2 & 13.0 \\
MPI synchronization  &  780.0 &  12.5 &  1.2 &  5.2 \\
    Other                &   13.5 &   4.5 &  4.3 &  6.5 \\
    \midrule
    Total                & 11916.1 & 1147.6 & 70.8 & 77.9 \\
    Speedup vs.\ CPU baseline & 1 & 10.4$\times$ & 168$\times$ & 153$\times$ \\
    \bottomrule
  \end{tabular}
\end{table*}


The computation of the matrix elements is the 
largest component of the wall time, and increases rapidly with the number of particles.
Thus, the optimization of this part plays a key role in increasing the number of particles and the GPU acceleration is most effective as discussed in Subsect.~\ref{sec:soa}.
In this part, the MI300A is 10\% faster than the GH200, although the theoretical peak of its vector arithmetic logic unit (VALU) is 80\% higher.
Measuring the FLOP count with the hardware counters of the GH200, the FP64 throughput of the matrix-element computation reaches 16.4~TFLOP/s, 48\% of the theoretical peak of the VALU of the GH200.
The corresponding throughput on the MI300A reaches 18.0~TFLOP/s, namely 29\% of its VALU peak. 
Note that the tensor cores are not used on either system for the present code since the size of the matrix $A$ is too small.

The ``Eigensolver'' part is the wall time of solving the eigenvalue problem with the \texttt{dsygvd} subroutine, provided by cuSOLVER 13.0 
on the GH200 and by rocSOLVER 7.2.0 on the MI300A every 100 basis states to avoid the accumulation of the numerical errors. The comparison indicates a higher level of maturity of the cuSOLVER implementation.

\subsection{6-body system}
\label{sec:6body}

We briefly discuss the possibility of extending the calculation to the six-body system. 
The five-body system with $N_t=14400$ and $N_b=9600$ was computed in 2574 seconds with 4 nodes, namely 2.9 node$\times$hours on Miyabi-G (GH200). With the same $N_t$ and $N_b$, the computational cost scales as $O(N^4N!)$, and thus the six-body calculation is expected to be 12.4 times more expensive than the five-body case. 
As a preliminary benchmark, we measured the wall time for the six-body case with the same conditions, and found that it is only 10.2 times larger. 
Although it is difficult to predict the values of $N_t$ and $N_b$ required for the six-body calculation, even if both $N_t$ and $N_b$ are doubled, the computational cost is estimated to be 237 node$\times$hours, which is feasible with current computational resources.

\section{Summary}
\label{sec:summary}
We presented the algorithm and code developments of the SVM for the five-body $^4$He atomic system.
The present code was highly tuned with the aid of AI coding.
Employing the secular equation for low-lying eigenvalues significantly accelerates the SVM computation and makes a large number of basis states feasible, which is directly relevant to the capability of the SVM.
We achieved a 168-fold acceleration on the GH200 over the baseline CPU code, and a 16-fold acceleration over the tuned CPU code. The loop exchange, which enables the efficient use of SIMD on CPUs and of coalesced addressing on GPUs, has a significant impact.
This code makes 6-body calculations and beyond computationally feasible.
Since the whole code was developed by an AI agent, 
the present study shows the ability of AI to write a performant scientific code for current GPU architectures, while a working knowledge of the hardware is required.
In particular, the secular-equation solver for massively parallel computation and the Gu--Eisenstat prescription that preserves the orthogonality of the eigenvectors are known to be efficient, but implementing them is a burdensome task for physicists. 
The recent growth of AI coding enables us to implement them in a short time and to exploit the latest HPC environments efficiently.

As a future work, further improvements of the SVM algorithm are anticipated, such as the gradient-based optimization \cite{gradient_svm} and Bayesian optimization \cite{aoyama_bayes}.

\section*{Acknowledgment}

\ifanon\else
N. S., S. A., and Y. T.  acknowledge the KAKENHI budgets (25H01268) and 
the  JST ERATO project (JPMJER2304). 
N. S. thanks also the KAKENHI budgets (25K00995, 25K07294, 25K07330, 24H00239).
\fi
The numerical calculations were performed 
at the Miyabi \cite{miyabi} and Sirius \cite{sirius} supercomputers  under the MCRP program of Center for Computational Sciences, University of Tsukuba.
The coding was performed mainly by Anthropic Claude Opus and Fable. 
The draft, figure and tables were prepared with the aid of Claude Opus.

\vspace{12pt}

\end{document}